\documentclass[conference,10pt]{IEEEtran}

\usepackage[cmex10]{amsmath}
\usepackage{amssymb,mathtools}
\usepackage{algorithm,algorithmic}
\usepackage{array}
\usepackage{mdwmath}
\usepackage{mdwtab}
\usepackage{fixltx2e}
\usepackage{url}
\usepackage{cmap}
\usepackage{animate}
\usepackage{xspace}
\usepackage{xcolor}
\usepackage{citesort}

\newlength\figureheight
\newlength\figurewidth

\newtheorem{example}{Example}

\newtheorem{theorem}{Theorem}

\newcommand{\SSM}{SSM\xspace}
\newcommand{\CPFAS}{CPF-AS\xspace}

\newcommand{\Exp}[1]{\mathbb{E}\left[{#1}\right]}
\newcommand{\Expb}[2]{\mathbb{E}_{#1}\left[{#2}\right]}
\newcommand{\Prb}[1]{\mathbb{P}\left({#1}\right)}
\newcommand{\Transp}{\mathsf{T}}
\newcommand{\M}{\mathcal{M}}
\newcommand{\ii}{{(i)}}
\newcommand{\ji}{{(j)}}
\newcommand{\ki}{{(k)}}
\newcommand{\N}{\mathcal{N}}
\newcommand{\s}{x}
\newcommand{\fx}{f_x}
\newcommand{\fu}{f_u}
\newcommand{\gx}{g_x}
\newcommand{\gu}{g_u}
\newcommand{\out}{y}
\newcommand{\anc}{a}
\newcommand{\w}{w}

\newcommand{\epdf}[1]{\mathcal{N}\left({#1}\mid \out_t,R\right)}
\newcommand{\qpdf}[2]{\mathcal{N}\left({#1}\mid {#2},Q\right)}
\newcommand{\qdist}[1]{\mathcal{N}\left({#1},Q\right)}
\newcommand{\setX}{\mathsf{X}}

\newcommand{\R}{\mathbb{R}}
\newcommand{\saemQ}{\mathbb{Q}}
\newcommand{\phivec}{{\bar{\varphi}}}

\begin{document}
\title{Nonlinear State Space Model Identification\\Using a Regularized Basis Function Expansion}

\author{\IEEEauthorblockN{Andreas Svensson$^{\star}$, Thomas B. Sch\"{o}n$^{\star}$, Arno Solin$^{\star\star}$, Simo S\"{a}rkk\"{a}$^{\star\star\star}$}\vspace{0.5em}
	\IEEEauthorblockA{\small \it $^\star$Department of Information Technology, Uppsala University, Sweden. E-mail:
		{\{andreas.svensson, thomas.schon\}@it.uu.se}\\
		$^{\star\star}$Department of Neuroscience and Biomedical Engineering, Aalto University, Finland. E-mail:
		{arno.solin@aalto.fi}\\
		$^{\star\star\star}$Department of Electrical Engineering and Automation, Aalto University, Finland. E-mail:
		{simo.sarkka@aalto.fi}%
	}}

\maketitle

\begin{abstract}
  This paper is concerned with black-box identification of nonlinear
  state space models. By using a basis function expansion within the
  state space model, we obtain a flexible structure. The model is
  identified using an expectation maximization approach, where the
  states and the parameters are updated iteratively in such a way that
  a maximum likelihood estimate is obtained. We use recent particle
  methods with sound theoretical properties to infer the states,
  whereas the model parameters can be updated using closed-form
  expressions by exploiting the fact that our model is linear in the
  parameters. Not to over-fit the flexible model to the data, we also
  propose a regularization scheme without increasing the computational
  burden. Importantly, this opens up for systematic use of
  regularization in nonlinear state space models. We conclude by
  evaluating our proposed approach on one simulation example and two
  real-data problems.
\end{abstract}
\IEEEpeerreviewmaketitle

\section{Introduction}
Modeling of nonlinear dynamical systems is a well-studied problem
within many areas, including system identification
\cite{Ljung:1999,Billings:2013} and time series analysis
\cite{Tsay:2010}. This paper is concerned with black-box
identification of nonlinear state space models (SSMs), where the
nonlinearities are represented using a basis function expansion. By
applying the recent method \cite{Lindsten:2013} based on sequential
Monte Carlo (SMC) and Expectation Maximization (EM), we can
efficiently utilize the linear-in-its-parameters and
nonlinear-in-the-state properties of our proposed model (see, e.g.,
\cite{KSS:2014} for discussion on such models).

The use of basis function expansions is a well-established approach
within system identification, often used to identify transfer 
functions \cite{HVW:2005}. An early approach to use
basis function expansions also for nonlinear system identification is
found in \cite{KFM:1978}, whereas \cite{Pawlak:1991,Greblicki:1994} are
using it to identify static nonlinearities in Hammerstein and Wiener
systems, respectively. The idea to combine a basis function expansion
and EM to identify nonlinear SSMs dates back---to the best of our
knowledge---to \cite{GR:1999}. A recent work along the same line is
\cite{PLS+:2010}, replacing EM with nonlinear optimization.

To avoid over-fitting to data, one remedy is to limit the number of terms in the series
expansion, as discussed by \cite{Pawlak:1991}. Another alternative is regularization \cite{COL:2012,CL:2015}, or equivalently, a prior on the basis function weights. The work by \cite{SSS+:2015} is addressing the
special case of a Gaussian process prior.

In particular, we consider identification of a nonlinear \SSM
\vspace{-1em}
\begin{subequations}\label{eq:ss}
\begin{align}
	x_{t+1} &= f_x(x_t) + f_u(u_t) + w_t\label{eq:ssa},\\
	y_t &= g_x(x_t) + g_u(u_t) + e_t,\label{eq:ssb}
	\end{align}
\end{subequations}
with $w_t$ and $e_t$ being Gaussian noise with zero mean and
$\Exp{w_t w_t^\Transp} = Q$, $\Exp{e_t e_t^\Transp} = R$, and 
$x_t \in \R^{n_x}$. By making use of input-output data $(u_{1:T},y_{1:T})$
and prior knowledge of $n_x$, our goal is to identify a model
$\M \triangleq \{f_x, f_u, g_x, g_u, Q, R\}$ maximizing the
marginal likelihood $p(u_{1:T},y_{1:T}|\M)$. The approach can be
generalized to functions $f(x_t,u_t)$ and alike.

In practical situations, it may be too ambiguous to identify \emph{all} functions
in~\eqref{eq:ss}. It is also possible to adapt the proposed scheme to learn only parts of~\eqref{eq:ss}, as will be illustrated by Example~\ref{sec:hw}.

\section{Orthogonal basis function expansions in {\SSM}s}
We restrict ourselves to consider a compact set $\setX$ of $\R^{n_x}$, and assume there exists a set of orthogonal basis functions $\{\phi^\ki\}_{k=1}^\infty$ spanning the function space on $\setX$ such that
\begin{align}
f_x(x) = \sum_{k=1}^\infty \omega_x^\ki\phi^\ki(x) \approx \sum_{k=1}^m \omega_x^\ki\phi^\ki(x),~\forall x\in\setX, \label{eq:basisexp}
\end{align}
and similar for $f_u(\cdot)$, $g_x(\cdot)$ and $g_u(\cdot)$.

\begin{example}\label{bas:fourier}
  The Fourier basis $\phi^\ki(x) = e^{\frac{i\pi kx}{L}}$ spans the
  space of $L^2$ (i.e., square-integrable) functions on $\setX = [-L,L]$.
\end{example}

The truncated basis function expansion \eqref{eq:basisexp} suggests
the following approximation of~\eqref{eq:ssa}

\vspace{-2ex}
{\footnotesize
\begin{align}
	& x_{t+1} = \nonumber\\
	&\underbrace{\left[\begin{smallmatrix} \omega_{x,1}^{(1)} & \cdots & \omega_{x,1}^{(m)} \\ \vdots & & \vdots \\ \omega_{x,n_x}^{(1)} & \cdots & \omega_{x,n_x}^{(m)}  \end{smallmatrix}\right]}_{A}
	\underbrace{\left[\begin{smallmatrix} \phi^{(1)}_{\vphantom{1}}(x_t) \\ \vdots \\ \phi^{(m)}_{\vphantom{n_x}}(x_t) \end{smallmatrix}\right]}_{\phivec(x_t)} + 
	\underbrace{\left[\begin{smallmatrix} \omega_{u,1}^{(1)} & \cdots & \omega_{u,1}^{(m)} \\ \vdots & & \vdots \\ \omega_{u,n_x}^{(1)} & \cdots & \omega_{u,n_x}^{(m)}  \end{smallmatrix}\right]}_{B}
	\underbrace{\left[\begin{smallmatrix} \phi^{(1)}_{\vphantom{1}}(u_t) \\ \vdots \\ \phi^{(m)}_{\vphantom{n_x}}(u_t) \end{smallmatrix}\right]}_{\phivec(u_t)}
	+ w_t\label{eq:ss2a}
\end{align}} \\
and analogously for~\eqref{eq:ssb}. (Note that different basis
functions $\{\phi^\ki\}_{k=1}^\infty$ can be used for $f_x(\cdot)$ and
$f_u(\cdot)$, although this is not reflected in the notation.) More
compactly we have the following approximation of~\eqref{eq:ss}
\begin{subequations}\label{eq:lsss}
	\begin{align}
	x_{t+1} &= \begin{bmatrix} A & B\end{bmatrix} \begin{bmatrix}\phivec(x_t) \\ \phivec(u_t)\end{bmatrix} + w_t,\\
	y_{t} &= \begin{bmatrix} C & D\end{bmatrix} \begin{bmatrix}\phivec(x_t) \\ \phivec(u_t)\end{bmatrix} + e_t,
	\end{align}
\end{subequations}
which is linear in the parameters, but nonlinear in the states~$x_t$
and the inputs~$u_t$. This structure will be exploited when we derive
the maximum likelihood (ML) estimator.

\section{ML Identification of \SSM{s}}
The approximate model~\eqref{eq:lsss} allows us to cast the original
problem of identifying the model~$\M$ as the problem of identifying a
(large) number of parameters 
$\theta \triangleq \{A, B, C, D, Q, R\}$ in~\eqref{eq:lsss}. This
reformulation of the original nonparametric model as a parametric
model opens up for use of recent tools for identification of nonlinear
parametric models.

\subsection{Nonlinear system identification using PSAEM}
To identify the parameters $\theta$ in~\eqref{eq:lsss}, several
methods can be used. We will make use of the recent development
\cite{Lindsten:2013}, relying on a combination of stochastic
approximation EM \cite{DLM:1999} and a conditional particle filter
with ancestor sampling (\CPFAS, \cite{LJS:2014}). The \CPFAS can be
interpreted as a particle smoother formulated as a Markov chain Monte
Carlo (MCMC) method, and is provided in
Algorithm~\ref{alg:CPFAS}. Stochastic approximation EM is compatible
with MCMC-methods \cite{KL:2004}, which allows the \CPFAS to be
combined into particle stochastic approximation EM (PSAEM).

\begin{algorithm}[t]
	\footnotesize
	\caption{\footnotesize Conditional particle filter with ancestor sampling}
	\label{alg:CPFAS}
	\begin{algorithmic}[1]
		\REQUIRE Conditional trajectory $\s_{1:T}[k]$.
		\ENSURE Trajectory $\s_{1:T}[k+1]$ and particle system $\{\s_{1:T}^\ii,\w_T^\ii\}_{i=1}^N$.
		\STATE Draw $\s_1^\ii \sim p(\s_1)$ for $i = 1,\dots,N-1$.
		\STATE Set $\s_1^{(N)} = \s_1[k]$.
		\FOR{$t = 1, \dots, T$}
		\STATE Set $w_t^\ii = \epdf{\gx(\s_t^\ii)+\gu(u_t)}$.
		\STATE Draw $\anc_t^\ii$ with $\Prb{\anc_t^\ii=j} \propto w_{t}^\ji$ for $i = 1, \dots, N-1$.
		\STATE Draw $\s_{t+1}^\ii \sim \qdist{\fx(\s_t^{\anc_t^\ii})+\fu(u_t)}$ for $i = 1, \dots, N-1$.
		\STATE Set $\s_{t+1}^{(N)} = \s_{t+1}[k]$.
		\STATE Draw $\anc_t^{(N)}$ with\\$\Prb{\anc_t^{(N)} = j} \propto w_{t}^\ji\qpdf{\fx(\s_t^\ji)+\fu(u_t)}{\s_{t+1}^{(N)}}$.
		\STATE Set $\s_{1:t}^i = x_{1:t}^{\anc_t^i}$ for $i = 1, \dots, N$.
		\ENDFOR
		\STATE Draw $J$ with $\Prb{J=i}\propto \w_T^\ii$ and set $\s_{1:T}[k+1] = \s_{1:T}^{(J)}$.
	\end{algorithmic}
\end{algorithm}

PSAEM will generate a sequence $\theta[1], \theta[2], \dots$
converging to a stationary point of $p_\theta(u_{1:T},y_{1:T})$. Algorithmically,
it starts from an arbitrary initial parameter $\theta[0]$ and a state
space trajectory $\s_{1:T}[0]$, and then iterates the following two
steps until convergence:
\begin{itemize}
	\item[(E)] run \CPFAS (Algorithm~\ref{alg:CPFAS}) with
          $\theta[k-1]$ as model to obtain
          $\{\s_{1:T}^\ii,\w_T^\ii\}_{i=1}^N$,
	\item[(M)] update the parameters $\theta[k-1] \mapsto
          \theta[k]$ as the maximizing argument to an auxiliary
          function $\saemQ_{k}(\theta)$ (detailed below).
\end{itemize}

The M-step in the PSAEM algorithm amounts to maximizing the auxiliary function

\vspace{-1.5em}
{\footnotesize
\begin{align}
\saemQ_k(\theta) = &(1-\gamma_k)\saemQ_{[k-1]}(\theta) + \gamma_k\left(w_T^\ji\sum_{j=1}^N \log p(x_{1:T}^\ji,y_{1:T}|\theta)\right)\label{eq:saemQ}
\end{align}}\vspace{-0.3em}\\
with $\{\gamma_k\}_{k\geq1}$ being a sequence fulfilling $\gamma_1 = 1$, \mbox{$\gamma_{k+1} < \gamma_k$}, $\sum_k \gamma_k = \infty$ and $\sum_k \gamma_k^2 < \infty$.
Due to the structure of our problem there is a closed-form solution
(see Theorem~\ref{thm:1}) 
available for the problem $\theta[k] = \arg\max_\theta
\saemQ_k(\theta)$. Here, 
$\Expb{\theta}{\cdot}$ denotes expectation under the model $\theta$.

\begin{theorem}[Maximizing $\saemQ$]\label{thm:1}
	Consider a model on the form
	\begin{align}
	\zeta_t = \Gamma z_t + v_t,
	\end{align}
	with $\Exp{v_t^\Transp v_t} = \Pi$ (cf. \eqref{eq:lsss}: $\zeta_t = x_{t+1}$, $\Gamma = [A~B]$, \mbox{$\Pi = Q$}, $z_t^\Transp = [\phivec(x_t)^\Transp \phivec(u_t)^\Transp]$). Further, $\zeta_t, z_t$ are given and $\theta \triangleq \{\Gamma,\Pi\}$ is to maximize. Then $\theta[k] = \arg\max_\theta \saemQ_k(\theta)$ is
	\begin{subequations}\label{eq:PSAEM_max}
		\begin{align}
		\Gamma[k] &= \Psi[k]\Sigma[k]^{-1},\label{eq:gammas}\\
		\Pi[k] &= \Phi[k] - \Psi[k]\Sigma[k]^{-1}\Psi[k],\label{eq:pis}
		\end{align}
		where
		\small
		\begin{align}
		\Phi[k] &= (1-\gamma_k)\Phi[{k-1}] +\tfrac{1}{T}\Sigma_{t=1}^T \Expb{\theta[k-1]}{\zeta_{t}^\Transp \zeta_{t}\mid y_{1:T}},\label{eq:PhiPSAEM}\\
		\Psi[k] &= (1-\gamma_k)\Psi[{k-1}] +\tfrac{1}{T}\Sigma_{t=1}^T\Expb{\theta[k-1]}{\zeta_{t}^\Transp z_t\mid y_{1:T}},\label{eq:PsiPSAEM}\\
		\Sigma[k] &= (1-\gamma_k)\Sigma[{k-1}] +\tfrac{1}{T}\Sigma_{t=1}^T \Expb{\theta[k-1]}{z_t^\Transp z_t\mid y_{1:T}}.\label{eq:SigmaPSAEM}
		\end{align}
	\end{subequations}
In case $\zeta_t$ or $z_t$ are given by a particle system $\{z_{1:T}^\ii,w_{T}^\ii\}_{i=1}^N$, the expectations are computed as
\begin{align}
&\Expb{\theta[k-1]}{z_{t}^\Transp z_{t}\mid y_{1:T}} = \int z_{t}^\Transp z_{t} p_{\theta[k-1]}({z_{1:T} \mid y_{1:T}})\,dz_t \nonumber\\&\approx \sum_{i=1}^N w_{T}^\ii{z_{t}^\ii}^\Transp z_{t}^\ii. \label{eq:Eapprox}
\end{align}
\end{theorem}

\textit{Proof of Theorem~\ref{thm:1}:} The proof is omitted due to space limitation, but follows from \cite{Lindsten:2013,GN:2005,DLM:1999}.\hfill$\square$

The identification procedure is summarized in Algorithm~\ref{alg:PSAEM_id}.

\begin{algorithm}[t!]
	\footnotesize
	\caption{\footnotesize ML identification of nonlinear \SSM using PSAEM}
	\label{alg:PSAEM_id}
	\begin{algorithmic}[1]
		\STATE Initialize $\theta{[0]}$ arbitrarily.
		\STATE Set $x_{1:T}[0]$ arbitrarily (conditional trajectory in \CPFAS).
		\FOR{$k \geq 1$}
		\STATE Run Algorithm~\ref{alg:CPFAS} with conditional trajectory $x_{1:T}[k-1]$ and $f_x(x) = A{[k-1]}\phivec(x)$, $f_u(x) = B{[k-1]}\phivec(x)$, etc.
		\STATE Compute $A{[k]}$, $B{[k]}$, etc. according to \eqref{eq:PSAEM_max}.
		\ENDFOR 
	\end{algorithmic}
\end{algorithm}

\subsection{Regularization}
The model \eqref{eq:lsss} is very flexible. When it
comes to identification this flexibility can cause problems, such as 
overly complex models over-fitted to the data or the lack of unique
solutions.

Regularization is one remedy to avoid such problems~\cite{COL:2012}. Intuitively, regularization amounts to solving the problem $\arg\max_\theta p_\theta(u_{1:T},y_{1:T})$ under the additional constraint of `keeping $\theta$ as small as possible'. We will extend our approach to incorporate $L^2$ regularization, or equivalently, assigning Gaussian priors to the weights $\omega$.

By assigning a zero-mean Gaussian prior with precision matrix $P$ to
the basis function weights, that is,
$p([\omega^{(1)}~\cdots~\omega^{(m)}]) \sim \N(0,P^{-1})$, the
developments in~\cite{WSL+:2012} can be used to derive the alternative version of \eqref{eq:gammas}-\eqref{eq:pis} for regularized identification
\begin{subequations}\label{eq:map}
	\begin{align}
	\Gamma[k] &= \Psi[k]\left(\Sigma[k]+\tfrac{1}{T}P\right)^{-1},\label{eq:gammasa}\\
	\Pi[k] &= \Phi[k] - \Psi[k]\left(\Sigma[k]+\tfrac{1}{T}P\right)^{-1}\Psi[k].\label{eq:pisa}
	\end{align}
\end{subequations}
It is clear that the use of a prior with infinite variance ($P = 0$) retrieves the non-regularized identification algorithm.

A natural question is indeed how to choose the prior precision $P$. As
stated by \cite{COL:2012}, the optimal choice (in terms of mean square
error) is
$P^{-1}_{\mathrm{opt}} =
\Exp{[\omega^{(1)}~\cdots~\omega^{(m)}]^\Transp[\omega^{(1)}~\cdots~\omega^{(m)}]}$,
if we think of $\omega^{(1)}, \dots, \omega^{(m)}$ as being random
variables. As an example, with the natural assumption of $f_x(\cdot)$
being smooth, the diagonal elements of $P$ should be larger with
increasing order of the Fourier basis functions. The special case of
assuming $f_x(\cdot)$ to be a sample from a Gaussian process is
addressed by~\cite{SSS+:2015}.

Other regularization schemes, such as $L^1$, are possible but will not result in closed-form expressions such 
as~\eqref{eq:map}.

\begin{figure}[t!]
	\centering
	\includegraphics{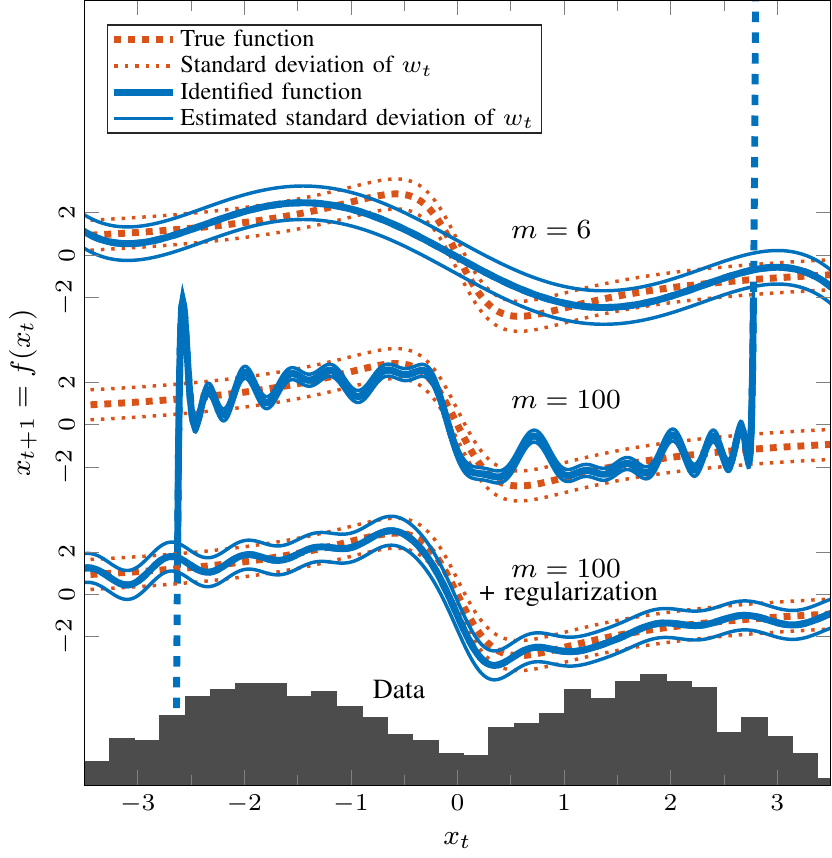}
	\vspace*{-1em}
	\caption{The first example, with three different settings:
          $m=6$ basis functions (top), $m = 100$ basis functions
          (middle) and $m = 100$ basis functions with regularization
          (bottom). The model with $m=6$ is not flexible enough to
          describe the `steep' part of $f$, but results in a sensible, albeit not perfect, model. The second model is very flexible with its $101$ parameters, and becomes a typical case of over-fitting to the data points (cf. the distribution of the data at the very bottom), causing numerical problems and a useless model. The regularization in the third case is a clear remedy to this problem, still maintaining the high flexibility of the model.}
	\vspace*{-.5em}
	\label{fig:simex}
\end{figure}

\subsection{Computational aspects}
Let $N$ denote the number of particles in the \CPFAS, $m$ the numer of
terms used in the basis function expansion, $T$ the number of data
points and $K$ the numer of iterations used in
Algorithm~\ref{alg:PSAEM_id}. The computational load is then 
$\mathcal{O}(mTKN) + \mathcal{O}(m^3)$. In practice, $N$ and $m$ can
be chosen fairly small (e.g., $N = 5$ and $m = 10$ for a {1D} model).

\subsection{Convergence}
%
The convergence properties of PSAEM are not yet fully understood, but
it can under certain assumptions be shown to converge to a stationary point
of $p_\theta(u_{1:T},y_{1:T})$ by \cite[Theorem 1]{KL:2004}. We have not
experienced practical problems with the convergence,
although it is sensitive to initialization when the dimension of
$\theta$ is large (e.g., $1\thinspace 000$ parameters).

\section{Numerical examples}
We demonstrate our proposed method on a series of numerical
examples. The source code is available via the web site of the first
author.

\subsection{Simulated example}
As a first simple numerical example, consider an autonomous system (i.e., no $u_t$) defined by
\begin{align}
x_{t+1} = \frac{-10x_t}{1+3x_t^2} + w_t, \qquad
y_t = x_t + e_t,
\end{align}
where $w_t \sim \mathcal{N}(0,0.1)$ and $e_t\sim \N(0,0.5)$. We
identify $f(\cdot)$ and $Q$ from $T = 1\thinspace 000$ simulated
measurements $y_{1:T}$, while assuming $g(\cdot)$ and $R$ to be
known. We consider three different settings with $m = 6$ basis
functions, $m = 100$ basis functions and $m = 100$ basis functions
with regularization, respectively, all using the Fourier basis. To
encode the a~priori assumption of $f(\cdot)$ being a smooth function, we
choose the regularization as a Gaussian prior of $w_k$ with standard
deviation inversely proportional to $k$. The results are shown in
Figure~\ref{fig:simex}, where the over-fitting problem for $m = 100$,
and how regularization helps, is apparent.

\begin{table}[t]
\renewcommand{\arraystretch}{1.0}
\caption{Results for the Hammerstein-Wiener Benchmark}
\vspace{-1em}
\label{tab:HW}
\centering
	\begin{tabular}{r|l}
		\multicolumn{2}{l}{~} \\ 
		\multicolumn{2}{l}{\emph{Experiment with $T = 2\,000$}} \\ \hline
		Mean simulation error & 0.0005 V \\
		Standard deviation of simulation error & 0.020 V \\
		RMS simulation error & 0.020 V \\
		Run time & 13 min \\
		\hline
	\end{tabular}
\end{table}

\subsection{Hammerstein-Wiener benchmark}\label{sec:hw}
To illustrate how to adapt our approach to problems with a given
structure, we apply it to the real-data Hammerstein-Wiener system
identification benchmark by~\cite{SSL:2009}. We will use a subset 
with $2\thinspace 000$ data points from the original data set for
estimation. Based on the domain knowledge provided by~\cite{SSL:2009}
(two third order linear systems in a cascade with a static 
nonlinearity between), we identify a model with the structure
\begin{subequations}
	\begin{align}
		\left[\begin{smallmatrix} x^1_{t+1} \\ x^2_{t+1} \\
                    x^3_{t+1} \end{smallmatrix}\right] &=
                                                         A_1\left[\begin{smallmatrix}
                                                             x^1_{t} \\
                                                             x^2_{t} \\
                                                             x^3_{t} \end{smallmatrix}\right]
          + Bu_t, \\
		\left[\begin{smallmatrix} x^4_{t+1} \\ x^5_{t+1} \\ x^6_{t+1} \end{smallmatrix}\right] &= A_2 \left[\begin{smallmatrix} x^4_{t} \\ x^5_{t} \\ x^6_{t} \end{smallmatrix}\right] + 
		\left[\begin{smallmatrix} \Sigma_k \omega^\ki\phi^\ki(x_t^3) \\ \phantom{x^3_{t}}0\phantom{x^3_{t}} \\ \phantom{x^3_{t}}0\phantom{x^3_{t}} \end{smallmatrix}\right], \\
		\begin{smallmatrix} y_t \end{smallmatrix} &= C\left[\begin{smallmatrix} x^4_{t} & x^5_{t} & x^6_{t} \end{smallmatrix}\right],
	\end{align}
\end{subequations}
where the superindex on the state denotes a particular component of
the state vector. Furthermore, we have omitted all noise terms for
notational brevity. There is only one nonlinear function, but
the linear parts can be seen as the special case where 
$\{\phi^\ki(x)\}_{k=1}^m = \{x\}$, which can directly be incorporated
into the presented framework.

We present the results in Table~\ref{tab:HW} (all metrics are with respect to the evaluation data from the original data set). We refer to \cite{HRR:2012} for a thorough evaluation of alternative methods.

\subsection{Nonlinear MIMO model}
We now consider black-box identification of a nonlinear MIMO system,
used as an example by the System Identification Toolbox
\cite{MathWorks:2015a}. The data consists of $188$ measurements from a
motorized camera. The input vector contains $6$~variables:
$3$~translational velocity components and $3$~rotational velocity
components. The output vector contains two variables: the position in
the image of a fixed (in 3D space) point.

\begin{figure}[t]
	\vspace*{-0.5em}
	\centering
	\includegraphics{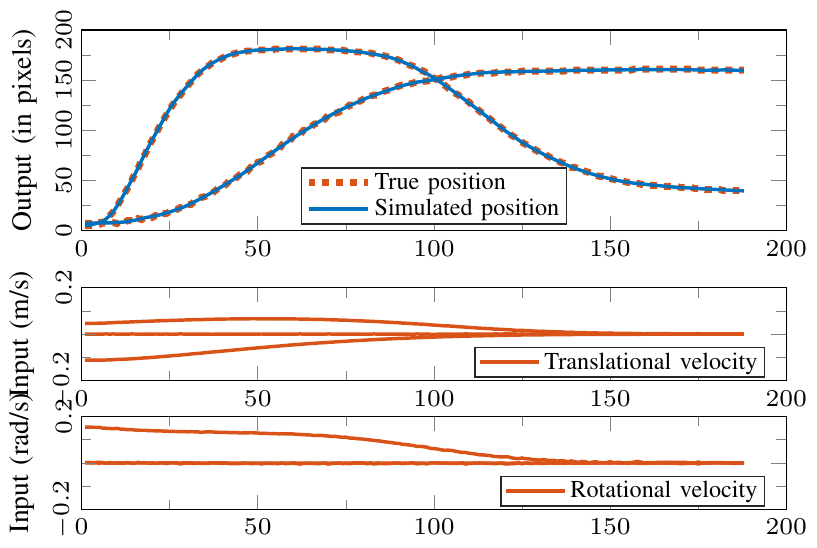}
	\vspace*{-1em}
	\caption{Modeling a motorized camera with 6 inputs and 2 outputs.}
	\vspace*{-.5em}
	\label{fig:camera}
\end{figure}

We model this using a two-dimensional state space and a known linear
measurement function $y_t = I_2 x_t + e_t$, and a simple
regularization, with a prior precision proportional to the order of
the basis functions. The results are displayed in
Figure~\ref{fig:camera}, where it can be seen that the simulated
output follows the true output very closely. The resulting RMSE is
$0.52$ pixels, whereas the best results reported by
\cite{MathWorks:2015a} for a nonlinear ARX (using a wavelet network)
is an RMSE of $2.22$ pixels, and $2.13$ for a rather complex Hammerstein
model.

\section{Conclusions and further work}
We have presented a model and 
algorithm for black box identification of nonlinear state-space models~\eqref{eq:ss}. Regularization is a key for its use in practice, providing a systematic approach to tune the
model complexity.

Within linear system identification it has recently 
\cite{COL:2012,CL:2015} been realized that the use of regularization
with carefully selected Gaussian process priors can enforce system
properties such as smoothness and stability. Our model construction
opens up for
similar developments also for nonlinear systems and a more systematic
approach to designing the regularization priors (e.g, akin to the
Gaussian process case \cite{SSS+:2015}) constitutes future
work. Besides this, some other topics for further investigation
would be the use of alternative basis functions (e.g., wavelets and Legendre
functions), the initialization of
Algorithm~\ref{alg:PSAEM_id}, and to replace the
particle filter with a smoother based on basis function expansions
akin to~\cite{RM:2014}.

\section*{Acknowledgment}

\footnotesize
This work was supported by the Academy of Finland (Projects 266940, 273475) and the Swedish Research Council (VR) (Project \emph{Probabilistic modeling of dynamical systems}, 621-2013-5524). We would also like to thank Dr.\ Olov Ros\'{e}n and Dr.\ Tianshi Chen for fruitful discussions on basis function expansions and regularization, respectively.

\bibliographystyle{IEEEtran}
\bibliography{IEEEabrv,references}

\end{document}